\documentclass[a4paper,11pt]{article}
\pdfoutput=1 
\usepackage{jcappub} 
\usepackage[normalem]{ulem}      
\usepackage[T1]{fontenc} 
\usepackage{bigints}
\usepackage{aas_macros}

\newcommand{\FermiLAT}{\textit{Fermi}-LAT }

\newcommand{\beq}{\begin{equation}}
\newcommand{\eeq}{\end{equation}}

\newcommand{\src}{4FGL J0616.2-0653}
\newcommand{\g}{$\gamma$}

\newcommand\orcid[1]{\href{https://orcid.org/#1}{$\!$\includegraphics[scale=0.0045]{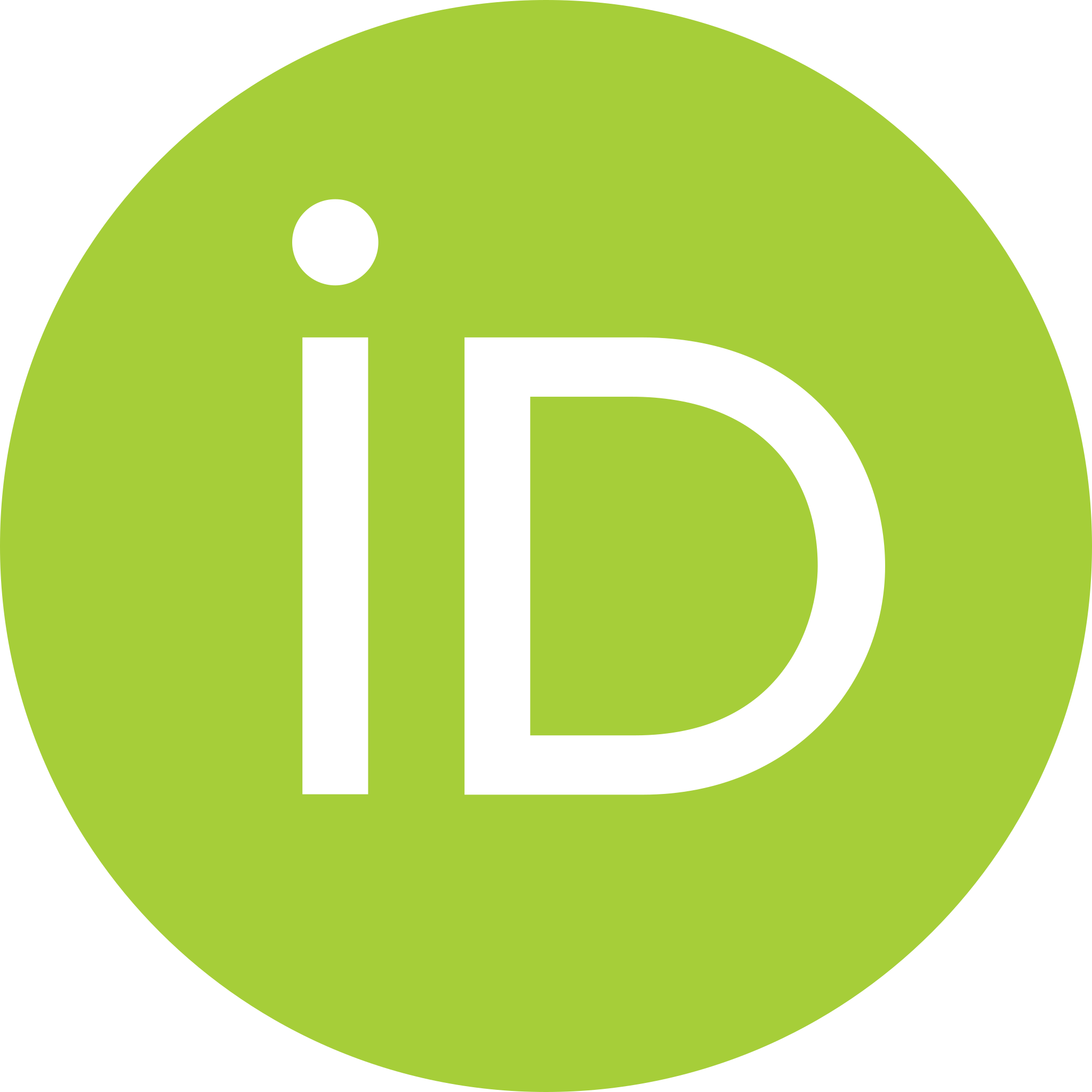} $\!\!$}}

\usepackage[left]{lineno}
\usepackage{graphicx}
\usepackage{dcolumn}
\usepackage{bm}
\usepackage{hyperref}
\usepackage{color}
\usepackage{adjustbox}
\usepackage{longtable}
\usepackage{multirow}
\usepackage{float}
\usepackage{subcaption}

\usepackage[dvipsnames]{xcolor}

\definecolor{km3net_blue}{HTML}{4479ff}

\hypersetup{
    colorlinks   = true,
    citecolor    = km3net_blue,
    urlcolor     = km3net_blue,
    linkcolor    = km3net_blue
}

\title{\boldmath Looking for the $\gamma$-Ray Cascades of the KM3-230213A Neutrino Source}

\author[a]{Milena Crnogor\v{c}evi\'{c} \orcid{0000-0002-7604-1779}}
\affiliation[a]{Stockholm University and the Oskar Klein Centre for Cosmoparticle Physics, Alba Nova, 10691 Stockholm, Sweden}

\author[b,c,d]{Carlos Blanco \orcid{0000-0002-2893-6594}}
\affiliation[b]{Department of Physics, The Pennsylvania State University, University Park, PA 16802, USA}
\affiliation[c]{Institute for Gravitation and the Cosmos, The Pennsylvania State University, University Park, PA 16802, USA}
\affiliation[d]{Institute for Computational and Data Sciences, The Pennsylvania State University, University Park, PA 16802, USA}

\author[a,e]{Tim Linden \orcid{0000-0001-9888-0971}}

\affiliation[e]{Erlangen Centre for Astroparticle Physics (ECAP), Friedrich-Alexander-Universität \\ Erlangen-Nürnberg, Nikolaus-Fiebiger-Str. 2,
91058 Erlangen, Germany}

\emailAdd{milena.crnogorcevic@fysik.su.se, carlosblanco@psu.edu, linden@fysik.su.se}

\abstract{
\noindent The extreme energy of the KM3-230213A event could transform our understanding of the most energetic sources in the Universe. However, it also reveals an inconsistency between the KM3NeT detection and strong IceCube constraints on the ultra-high energy neutrino flux. The most congruous explanation for the KM3NeT and IceCube data requires KM3-230213A to be produced by a (potentially transient) source fortuitously located in a region where the KM3NeT acceptance is maximized. In hadronic models of ultra-high-energy neutrino production, such a source would also produce a bright $\gamma$-ray signal, which would cascade to GeV--TeV energies due to interactions with extragalactic background light. We utilize the \texttt{$\gamma$-Cascade} package to model the spectrum, spatial extension, and time-delay of such a source, and scan a region surrounding the KM3NeT event to search for a consistent $\gamma$-ray signal. We find no convincing evidence for a comparable \textit{Fermi}-LAT source and place constraints on a combination of the source redshift and the intergalactic magnetic field strength between the source and Earth.  
}

\begin{document}
\maketitle
\flushbottom

\section{Introduction}
\label{sec:intro}

The unexpected detection of a signal by the \mbox{KM3NeT/ARCA} Collaboration that is consistent with a $\gtrsim$100~PeV neutrino has potentially opened a new window into the most extreme accelerators in the Universe~\cite{KM3NeT:2025npi}. If this event is confirmed to be an astrophysical neutrino, KM3-230213A 
stands out as the highest energy neutrino observed to date, raising questions regarding its astrophysical~\cite{KM3NeT:2025aps, KM3NeT:2025bxl, Dzhatdoev:2025sdi, Filipovic:2025ulm} or cosmological~\cite{KM3NeT:2025vut, Choi:2025hqt} origins, as well as potential multiwavelength counterparts. This event prompted a variety of theoretical interpretations involving beyond Standard Model physics~\cite{Brdar:2025azm}, including Lorentz invariance violation~\cite{Yang:2025kfr, Satunin:2025uui, Cattaneo:2025uxk} and dark matter decay or interactions~\cite{Borah:2025igh, Kohri:2025bsn, Barman:2025hoz, Jho:2025gaf, Jiang:2025blz, Narita:2025udw, Murase:2025uwv}. Scenarios involving evaporating primordial black holes have also been explored~\cite{Boccia:2025hpm, Klipfel:2025jql}. The tension between the KM3NeT signal and the lack of similar detections by the larger IceCube detector motivates scenarios where the KM3NeT signal is produced by a transient source fortuitously located in the optimal region for KM3NeT detection~\cite{Li:2025tqf, Neronov:2025jfj}.

If the neutrino flux is generated via the hadronic interactions of ultra-high energy protons or nuclei, there should also be an accompanying flux of high-energy $\gamma$-rays~\cite{Salamon:1997ac, 2010ApJ...710.1530V, Murase:2012df, Berezinsky:2016feh}. These $\gamma$-rays can initiate electromagnetic cascades on the extragalactic background light (EBL), shifting the bulk of their energy into the GeV--TeV band, where they could be observable by the \textit{Fermi} Large Area Telescope (LAT). In the case of steady-state sources, these cascades could produce a diffuse $\gamma$-ray halo surrounding the source, the size and spectrum of which would depend sensitively on the intergalactic magnetic field (IGMF) strength. In the case of transient sources, the spectral evolution of the cascade is more difficult to model; however, the spatial evolution is simpler, as the $\gamma$-ray cascade produced over short time delays has a spatial extension much smaller than the resolution of current $\gamma$-ray instruments. 

In this paper, we generate $\gamma$-ray cascades corresponding to the KM3NeT neutrino signature using the \texttt{$\gamma$-Cascade} code~\cite{Blanco:2018bbf, Capanema:2024nwe}, and then perform a time-resolved search for GeV $\gamma$-ray emission in the \textit{Fermi}-LAT energy band. As the angular uncertainty of the KM3NeT event spans several degrees, we conduct a systematic scan over a grid of locations to search for coincident $\gamma$-ray signals. We find no convincing evidence for any statistically significant $\gamma$-ray excess in any temporal or angular bin. Because the relative strength of the neutrino and GeV $\gamma$-ray signals depends sensitively on the source distance and the IGMF strength, we translate our null detections into constraints on these properties, operating under the assumption that the KM3NeT source is either a transient or steady-state neutrino emitter. 

In addition to our primary search, we identify a previously undetected \textit{Fermi}-LAT source approximately $2^\circ$ from the best-fit position of the KM3-230213A event, inside the 99\% KM3NeT localization ellipse. While this source generally has a spectrum softer than that expected for the GeV cascades of a transient source, we do note a one-year period \emph{prior} to the KM3-230213A neutrino event (2010--11) where the $\gamma$-ray flux has a particularly hard spectrum, and may potentially be interpreted as the $\gamma$-ray cascade from a previous outburst. However, our analysis indicates that the magnetic field strengths required to produce such temporally localized $\gamma$ rays fall below lower limits on the magnetic field strength inferred from TeV blazar observations. Finally, we discuss alternative hypotheses, including the possibility that the KM3NeT source is a so-called ``hidden" neutrino accelerator.


\section{Methodology and Data Analysis} 
\label{sec:data}

\subsection{Electromagnetic Cascade Modeling}
We model the electromagnetic counterpart of the neutrino signal under the assumption that a monoenergetic hadronic transient event at energies of $\mathcal{O}(100\text{ PeV})$ gave rise to the detected neutrino. We set the source luminosity to the total neutrino energy flux as calculated by KM3NeT, \mbox{$E_\nu^2 \Phi_\nu = 5.8\times 10^{-8} \text{ GeV}\;\text{cm}^{-2}\;\text{s}^{-1}\;\text{sr}^{-1}$}, which allows us to normalize the injected $\gamma$-ray energy up to a constant that accounts for the uncertainty in the production mechanism. We adopt a model for the injected $\gamma$-ray luminosity as follows:

\begin{equation}
    E_\gamma^2 \frac{dN}{dE_{\gamma}}(z) = \frac{4\pi d_L^2(z)}{1+z}\times \eta \left(4E_\nu^2 \frac{dN(z)}{dE_\nu dA}\right)\bigg|_{E_\gamma = 2E_\nu (1+z)},
\end{equation}
where $\eta \sim \mathcal{O}(1)$ accounts for the uncertainty in the production mechanism (e.g. $\eta = 1$ for photo-pion production or $\eta = 1/2$ for proton-proton interactions), and $dN(z)/dE_\nu dA = 4\pi \Phi_\nu \tau_{\text{ARCA}}$ is the integrated differential number flux of neutrinos detected over an active observation time of KM3NeT/ARCA high-energy neutrino detector, $\tau_{\text{ARCA}}\approx 1\; \text{year}$. Furthermore, we assume that the \g-rays are produced with a spectral shape that is centered about 440 PeV and whose width is narrow enough such that essentially all the injected energy is reprocessed into cascade photons. Practically, we model this as a log-normal distribution with a variance (width) of up to a decade. We note that our results are not sensitive to the spectral width of the injected $\gamma$-rays as long as this width is less than $\sim$3 decades. At larger widths, there would be a significant number of unattenuated TeV $\gamma$-rays for distances below $\sim$10 Mpc. If the emission extends below 100 GeV, there would be an additional unattenuated $\gamma$-ray component which may be bright compared to the KM3NeT flux, and would typically strengthen the resulting constraints in this paper.

All the primary $\gamma$-rays in our model will be attenuated due to pair production with the background radiation field stemming from a combination of the cosmic microwave background (CMB) and the EBL. We adopt the EBL model from Saldana-Lopez
et al. (2021) \cite{Saldana-Lopez:2020qzx}. After $e^{+/-}$ (hereafter, collectively called electrons) pair produce, these electrons are able to inverse-Compton scatter with the same background radiation, producing \g-rays. This cascade cycle reprocesses the energy above $\sim\mathcal{O}(\text{TeV})$ into GeV photons that are visible to the \textit{Fermi}-LAT. We note that while the highest-energy electrons experience Klein-Nishina (KN) suppression of their cooling rates, the cascade quickly develops a population of lower-energy electrons ($\sim$1--100~TeV) for which Thomson scattering provides an accurate description of the energy losses \cite{Murase:2011yw}. We compute these cascades semi-analytically using \texttt{$\gamma$-Cascade} \cite{Blanco:2018bbf}. Figure~\ref{fig:main} shows a comparison of our cascade spectra with Monte Carlo results from the ELMAG package, demonstrating good agreement between the two approaches for different IGMF strengths.

As the primary \g-rays leave the source at comoving distance $d_s$, they will cascade and change their original direction of travel. This is caused by the motion of the electrons under the influence of the IGMF. Once electrons are generated via pair production, they experience an acceleration perpendicular to their velocity, which causes them to turn with a Larmor radius, $R_L = E_e/eB(z)$, where $B(z)  = B_0 (1+z)^{-\alpha}$ is the IGMF strength at redshift $z$. Here we take $\alpha = 4$ for IGMF evolution driven by a dynamo mechanism, as predicted by theoretical models (see e.g. \cite{Turner:1987bw, Durrer:2013pga, Pomakov:2022cem}). This treatment assumes that the IGMF coherence length $\lambda_B$ is much larger than the electron cooling length, allowing us to define a meaningful Larmor radius over the cascade development scale. This assumption is well-justified for typical IGMF parameters: we adopt coherence lengths $\lambda_B \sim 1$ Mpc, while inverse Compton cooling against the CMB occurs over much shorter distances of $\sim$10~kpc for 1~PeV electrons (including KN suppression) and $\sim$40 kpc for 10 TeV electrons in the Thomson regime. At 440~PeV, the KN-corrected cooling length is $\sim 0.3$~Mpc. Thus, $\lambda_B > d_{\rm cool}$ even at the earliest, KN-suppressed stage, and $\lambda_B \gg d_{\text{cool}}$ once the cascade develops into the TeV range that dominates the final deflection \cite{Murase:2011yw}. The total angle by which the electrons' trajectory becomes deflected is given by,
\begin{equation}
    \delta = c \int_{E_{e,0}}^{E_e} dE_e'   \left( \frac{dE'_e}{dt}\right)^{-1} \frac{1}{R_L},
\end{equation}
where the energy loss rate of the electrons is given by,
\begin{equation}
\label{eq:thompson}
    \frac{dE_e}{dt} = \frac{4}{3} c \sigma_T u_{\text{CMB}} \gamma_e^2,
\end{equation}
and where $u_{\text{CMB}}$ is the energy density of the background radiation field, $\sigma_T$ is the Thomson cross section, and $\gamma_e = E_e/m_e c^2$ is the gamma factor of the electrons. Note that while we show the Thomson limit in Eq.~\ref{eq:thompson} for simplicity, our \texttt{$\gamma$-Cascade} simulations include full Klein-Nishina corrections for inverse Compton scattering at all energies. We can evaluate the integral above in terms of the energy-loss distance of the electrons, $d_{\text{ICS}} = c E_e /(dE_e/dt)$,

\begin{align}
    \delta(E_e,z) &= c \int_{E_{e,0}}^{E_e} dE_e'   \left( \frac{dE'_e}{dt}\right)^{-1} \frac{1}{R_L} \approx \frac{d_{\text{ICS}} (E_e)}{2R_L (E_e)} \nonumber \\
     &= \frac{3 m_e^2 c^4 e B(z)}{8 \sigma_T u_{\text{CMB}}(z) E_e^2} \nonumber \\
     &= \frac{3 m_e^2 c^4 e B_0}{8 \sigma_T u_{\text{CMB},0} E_e^2 (1+z)^{4+\alpha}}.
\end{align}
where we have used the fact that the CMB number density scales as $(1+z)^3$. 

We further assume a small-angle approximation, where $\delta \ll \pi$, (i.e. $d_{\text{ICS}} \ll R_L$). As the cascade progresses, the mean free path of the photons grows rapidly below $E_\gamma \sim \mathcal{O} (\text{PeV}) $. At energies above the point where the mean free path reaches its minimum, the mean free path scales as $D_{\gamma \gamma} = \langle\bar{\sigma}_{\gamma \gamma}(E_\gamma) n_{\text{CMB}} \rangle^{-1} \propto E_\gamma$.  Therefore, one can think of the cascade as being composed of several cycles that extend out  from the source up to characteristic distances set by the mean free path of the second-to-last  pair-production step, {$\ell_{\rm pair} (z) \sim D_{\gamma\gamma} (E_\gamma^0, z)$, where $E_\gamma^0$ is the energy of that second-to-last photon (and the final-generation electron has energy $E_e =E_\gamma^0/2)$. After this point, the original direction of the cascading photons has been changed by $\delta$ from it's original trajectory. 

The ``last'' photon is the one that is detected on Earth by the observer coming from a direction $\theta_{\text{obs}}$ with respect to the line of sight to the source. The observed photon energy $E_\gamma$ from the final inverse-Compton scattering on the CMB is related to the last-generation electron energy by $E_\gamma \simeq 4/3 \left( E_e / m_e c^2 \right)^2 \epsilon_{\rm CMB}$, implying $E_e \propto \sqrt{E_\gamma}$ \cite{Fang:2025nzg}\footnote{This scaling underlies the $E_\gamma^{-1}$ dependence of $\theta_{\text{obs}}$ in Eq.~\ref{eq:theta}}. It has been shown in several studies that this angle, which manifests in the characteristic apparent angular extension of the source, can be approximated by the following,
\begin{align}
        \theta_{\text{obs}}(E_\gamma,z)&\approx \frac{D_{\gamma \gamma }({E_\gamma^0}, z)}{d_A} \delta(E_e = E_\gamma^0/2,z), \text{and} \\
    \label{eq:theta}
    \theta_{\text{obs}}(E_\gamma,z=0)&\approx 1'' \times \left(\frac{E_\gamma}{100\; \text{GeV}}\right)^{-1}\left(\frac{B_0}{10^{-16}\;  \text{G}}\right) \left(\frac{D_{\gamma\gamma}(E^0_\gamma,z)/d_A}{100}\right)^{-1},
\end{align}
where $d_A = d_c/(1+z)$ is the angular diameter distance. We approximate the profile of the photons arriving at earth by a Gaussian distribution $\mathcal{N}(\mu_\theta =0,\sigma_\theta = \theta_{\text{obs}})$ centered at $\mu_\theta = 0$ with a characteristic width of $\sigma_\theta = \theta_{\text{obs}}(E_\gamma,z)$.  For a given observed angle, the observed time delay due to the additional travel distance of the cascade is given by,
\begin{align}
    \tau_{\text{obs}}(E_\gamma,z) &\approx (1+z)\frac{D_{\gamma \gamma}(E_\gamma^0,z)}{2} \delta^2(E_\gamma^0/2,z) \nonumber \\
    &\approx 5\, \text{yrs}\times (1+z)^{-(7+2\alpha)} \left(\frac{D_{\gamma \gamma}(E_\gamma^0, z)}{10\; \text{Mpc}} \right) \left(\frac{B_0}{10^{-16}\; \text{G}} \right)^2  \left(\frac{E_\gamma}{100 \; \text{GeV}} \right)^{-2}.
    \label{eq:delay}
\end{align}

Notice that since $\tau_{\text{obs}}\propto \theta_{\text{obs}}^2 $, the time delay of the \g-rays is distributed as a chi-squared distribution with one degree of freedom, $\chi^2_1((\tau-\tau_s)/\tau_{\text{obs}})$, where $\tau_s$ is the light-travel time to the source and $\tau$ is the time at which an observation is made . This is a consequence of our simplified model where the photons' arrival is distributed as an angular Gaussian. A key result is that the total integrated flux of \g-rays collected after a total observation time $T_{\text{obs}}$ is given by the CDF of this chi-squared distribution,
\begin{align}
    E_\gamma^2 \frac{dN}{dE_{\gamma}}(z, E_\gamma,T_{\text{obs}}) &= E_\gamma^2 \frac{dN}{dE_{\gamma}}(z, E_\gamma,\infty)\times \text{CDF}\left[\chi^2_1(T_{\text{obs}}/\tau_{\text{obs}}(E_\gamma,z))\right] \nonumber \\
    &\approx E_\gamma^2 \frac{dN}{dE_{\gamma}}(z, E_\gamma,\infty) \times \sqrt{\frac{2}{\pi}\frac{T_{\text{obs}}}{ \tau_{\text{obs}}(E_\gamma,z)}}, \text{if } T_{\text{obs}} \ll \tau_{\text{obs}} \nonumber \\
    &\approx 0.36 \times (1+z)^{(7/2+\alpha)} \left( \frac{B_0}{10^{-16} \text{ G}}\right)^{-1} \left( \frac{E_\gamma}{100 \text{ GeV}}\right) \nonumber \\ & \; \;\;\;\;\; \times E_\gamma^{2} \frac{dN}{dE_{\gamma}}(z, E_\gamma,\infty)
    \label{eq:approx}
\end{align}
where $dN/dE_\gamma (T_{\text{obs}}\rightarrow\infty)$ is the cascade spectrum of a steady-state source normalized to the same luminosity. The second line is true only  in the limit where the observation time $T_{\text{obs}}$ is much smaller than $\tau_{\text{obs}}$. The last line is a rough scaling that is valid, e.g. when observing a source that is about 10 Mpc away for $T_{\text{obs}} = 1~\text{yr}$ if $E_\gamma\ll \mathcal{O}(10 \text{ TeV})$ \textit{and} $B_0\ll \mathcal{O}(10^{-14} \text{ G})$. We compute the inclusive cascade spectra using \texttt{$\gamma$-Cascade} and predict the finite-observation spectrum using our simplified model. We note that since $\theta_{obs}\ll 1^\text{o}$ for observation times $T_{\text{obs}}\sim \mathcal{O}( \text{yrs})$, the angular extension of the cascade will not be resolved within the PSF of \textit{Fermi}-LAT. Therefore, the only transient observable we consider is the afterglow ``echo'' of the electromagnetic cascades.   

\begin{figure}[H]
\centering
    \includegraphics[width=0.56\linewidth]{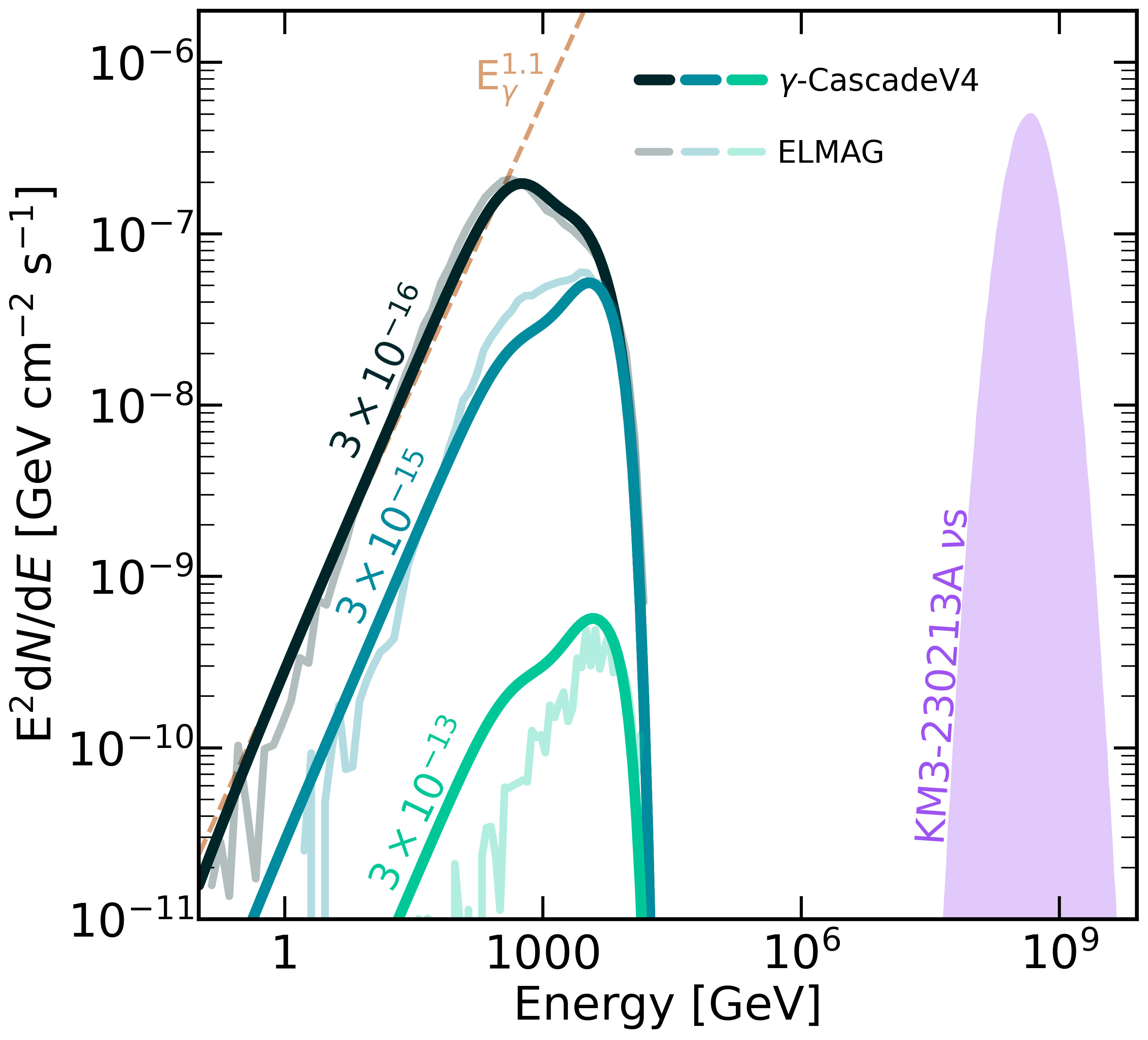}
    \caption{Comparison of visible cascades for $T_{\text{obs}} = 1$ year, $z=0.1$, and varying IGMF strengths, computed assuming IGMF evolution with $\alpha = 4$. Our results, computed using the semi-analytic \texttt{$\gamma$-CascadeV4} code, are shown in darker shades of teal-green curves, and are compared to the Monte Carlo-based \texttt{ELMAG} package in Ref.~\cite{Fang:2025nzg} shown in lighter shades. The orange dashed line represents the approximation given in Eq.~\ref{eq:approx}, applied to the universal spectrum of a fully developed electromagnetic cascade, $dN/dE_\gamma \propto E_\gamma^{-1.9}$~\cite{Berezinsky:2016feh}, from which we predict an energy spectrum as a power-law with spectral index of $\Gamma \sim$ 0.9. The expected KM3NeT neutrino spectrum under the assumption of photo-pion production is shown in pink.}
    \label{fig:main}
\end{figure}
\subsection{\emph{Fermi}-LAT \g-ray Analysis}
\label{subsec:fermi_analysis}

The \emph{Fermi} Large Area Telescope (LAT) is a pair-conversion $\gamma$-ray instrument onboard the \emph{Fermi} Gamma-ray Space Telescope, sensitive to energies from $\sim$100~MeV up to a few hundred GeV. With a large field of view ($\sim$2.4~sr), the LAT surveys the entire sky approximately every 3 hours. The improved event reconstruction provided by the \texttt{Pass8} data set enhances both the angular resolution and effective area, making the LAT particularly well-suited for detecting and characterizing steady-state and transient \g-ray sources. 
Further details regarding the LAT design, capabilities, and performance can be found in Ref.~\cite{LATinstrument:2009}.

In this analysis, we utilize publicly available \FermiLAT data spanning nearly 17 years, from August 4, 2008, to February 1, 2025, obtained from the \emph{Fermi} Science Support Center (FSSC)\footnote{\url{https://fermi.gsfc.nasa.gov/ssc/data/}, accessed on March 7, 2025.}. We select events classified as \texttt{P8R3\_SOURCE} and incorporate both \texttt{FRONT} and \texttt{BACK} conversion types. We apply a zenith angle cut of $100^\circ$ to reduce Earth-limb contamination and employ spatial binning with a pixel size of $0.1^\circ$, using logarithmic energy binning with eight bins per decade between 100~MeV and 500~GeV.

We define the region-of-interest (RoI, which is defined differently from that of KM3NeT, as the term is typically used to describe the region used to fit the $\gamma$-ray background) for our $\gamma$-ray study as a $15^\circ$ radius centered on the reported best-fit neutrino position from KM3NeT \mbox{(RA = 94.3$^\circ$, Dec = $-7.8^\circ$).} To account for localization uncertainties in the neutrino position, we perform a systematic grid search spanning $\pm2^\circ$ around the KM3NeT coordinates in increments of $0.2^\circ$. This choice of spatial increment is informed by the typical correlation scale of Test Statistic (TS) values for faint LAT point sources~\cite{Bertoni:2016hoh}. Here, the TS quantifies the source detection significance and is defined as $\text{TS} = -2\log(\mathcal{L}{_\text{null}}/\mathcal{L}_{\text{alt}})$, where $\mathcal{L}_{\text{alt}}$ and $\mathcal{L}_{\text{null}}$ denote the likelihood values with and without the source, respectively. In this context, the TS value serves as a measure of the statistical significance of a source, and is approximately related to the detection significance in units of standard deviation, $\sigma$, by $\sqrt{\text{TS}}\approx\sigma$ \cite{wilks1938}. The analysis is conducted consistently across various temporal intervals---weekly, monthly, yearly, and over the full dataset---to investigate potential steady and transient emission scenarios. 

\begin{figure}[H]
    \centering
    \includegraphics[width=0.7\linewidth]{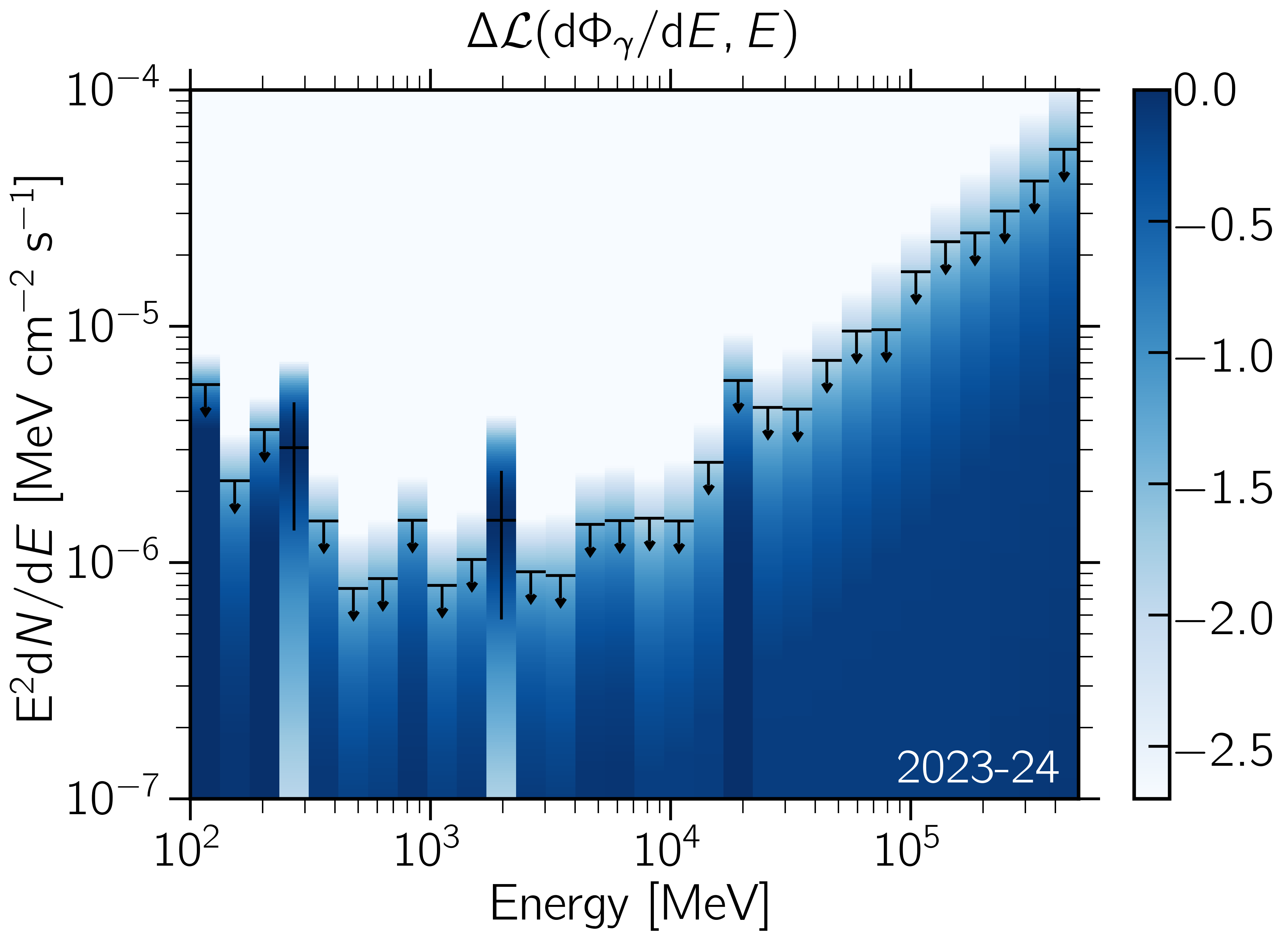}
    \caption{Spectral energy distribution (SED) and delta-log-likelihood ($\Delta \mathcal{L}$) profile for the region centered on KM3-230213A, covering 100~MeV to 500~GeV, during the 2023--24 time frame following the reported neutrino event. The black arrows represent 95\% CL flux upper limits, obtained under the assumption of a power-law spectrum with index $\Gamma=0.9$, appropriate for a one-year cascade timescale. The color scale indicates $\Delta \mathcal{L}$ as a function of flux and energy.}
    \label{fig:ul_sed}
\end{figure}

We model the diffuse background using the standard Galactic interstellar emission model (\texttt{gll\_iem\_v07})  and the isotropic diffuse emission model (\texttt{iso\_P8R3\_SOURCE\_V3}) provided by the LAT Collaboration\footnote{\url{https://fermi.gsfc.nasa.gov/ssc/data/access/lat/BackgroundModels.html}, accessed on March 7, 2025.}. Additionally, we include all known sources from the 4FGL-DR4 catalog within a $20^\circ$ radius around the RoI center to accurately constrain photon leakage into the \g-ray RoI \cite{Fermi-LAT:2022byn}. We process the data using the standard \texttt{FermiTools} (v.~2.2.0)\footnote{\url{https://fermi.gsfc.nasa.gov/ssc/data/analysis/software/}, accessed on March 7, 2025.} in conjunction with the open-source \texttt{fermipy} Python wrapper (v.~1.3.1) \cite{fermipy:2017}. 

First, we optimize and fit the RoI likelihood model. After the initial fit, we further optimize the model by freeing the normalization parameters of the diffuse background components as well as the parameters of nearby sources. Specifically, we free the normalization parameter for all sources within 5$^\circ$ of the RoI center. 

The spectral energy distribution (SED) for candidate sources is derived by independently fitting fluxes in each energy bin, assuming a power-law spectrum with an index $\Gamma=0.9$ (a choice justified by the hardness of the cascade spectrum shown in Fig.~\ref{fig:main} for a time delay of 1 year). We note, however, that choosing different $\gamma$-ray spectra ``inside'' each individual energy bin has a very small effect on our analysis, due to the small width of each bin. For energy bins lacking significant excess emission, we compute flux upper limits at a 95\% confidence level (CL) using the delta-log-likelihood method. An example of an SED curve upper limit assuming a power-law index of $\Gamma=0.9$ is shown in Fig.~\ref{fig:ul_sed}.

\begin{figure}[H]
    \centering
    \includegraphics[width=0.8\linewidth]{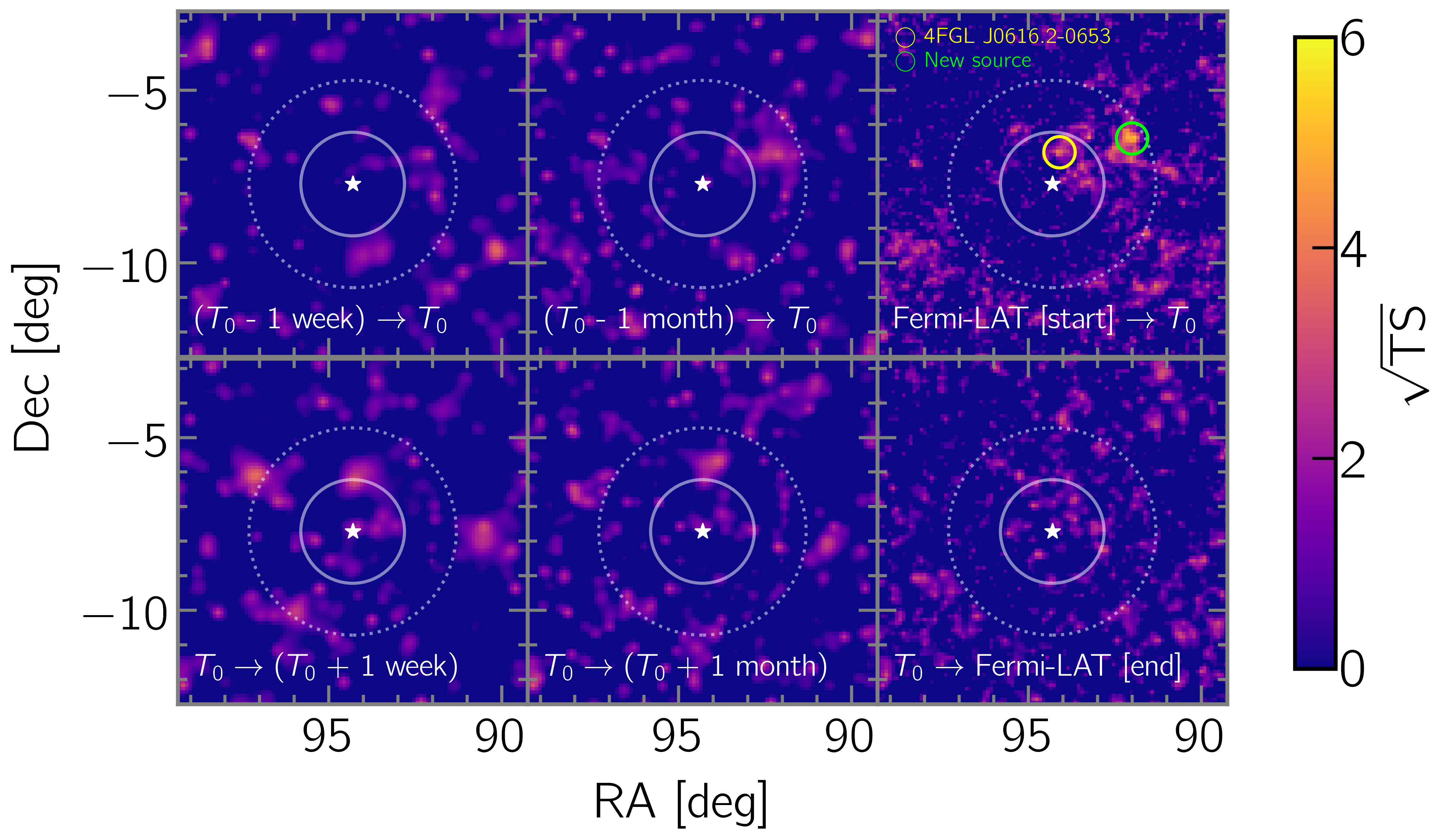}
    \caption{Time-resolved \g-ray residual maps from the \textit{Fermi}-LAT, showing intervals of one week (left), one month (center), and the total (right) \textit{Fermi}-LAT exposure relative to the KM3-230213A neutrino event time ($T_0$). The color scale represents $\sqrt{\mathrm{TS}}$, indicating the detection significance. The white star denotes the best-fit position of KM3-230213A. The solid white circle represents $1.5^\circ$ containment radius (68\% KM3NeT containment) while the dashed circle corresponds to $3^\circ$ (99\% KM3NeT containment). No significant transient or steady-state \g-ray emission is observed at the location of KM3-230213A. We note a new bright, unassociated excess observed in the upper-right region of the ``\FermiLAT [start] $\rightarrow$ $T_0$'' residual map (green circle), approximately $2^\circ$ northwest from the reported neutrino position. The known 4FGL J0616.2-0653 source is also shown in the same residual plot, indicated by the yellow circle.}
    \label{fig:panel1}
\end{figure}

\newpage
\begin{figure}[H]
    \centering
    \includegraphics[width=0.8\linewidth]{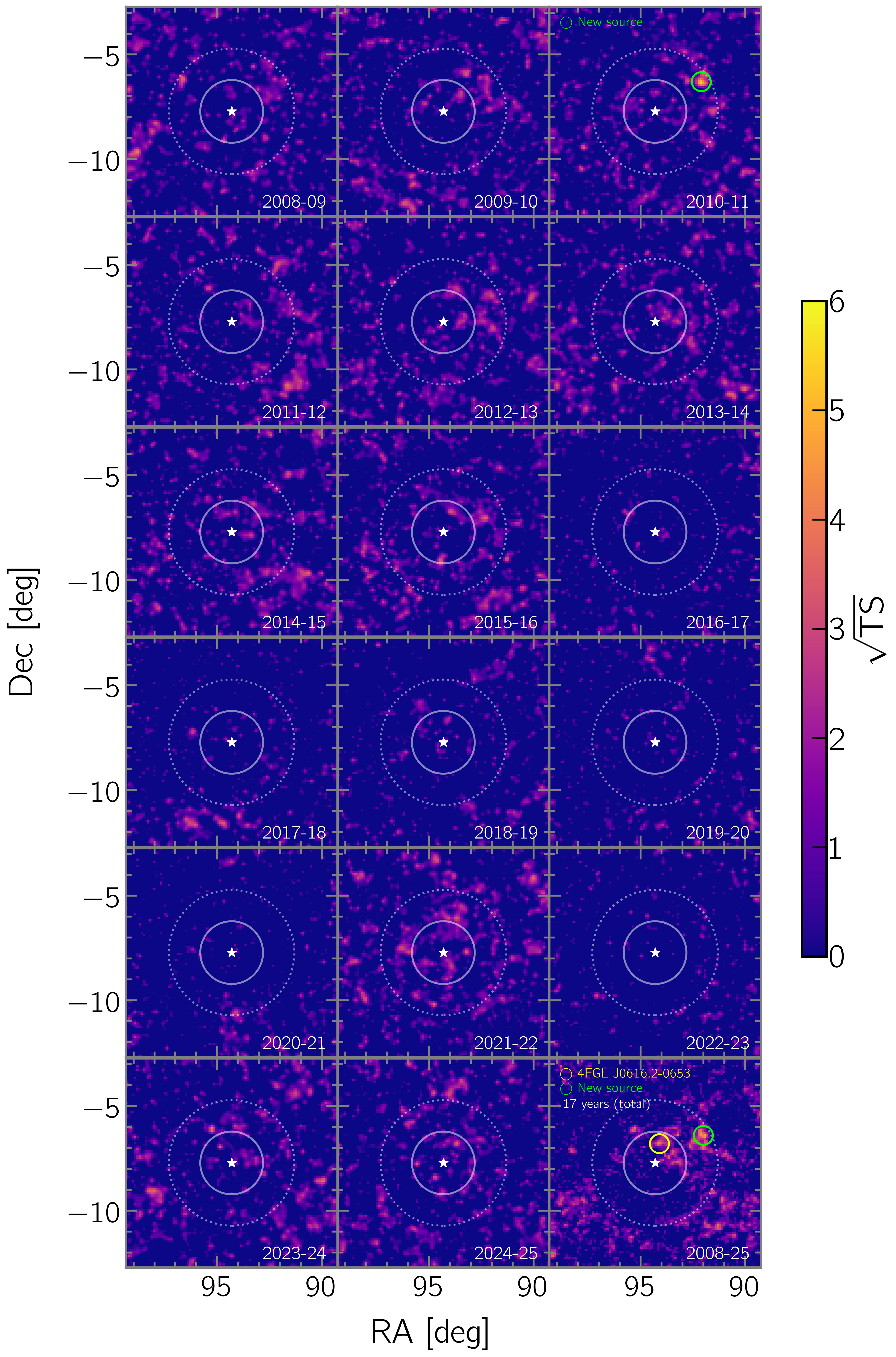}
    \vspace{-0.2cm}
    \caption{Analogous to Fig.~\ref{fig:panel1}, but showing \g-ray residual maps binned in one-year intervals defined relative to the KM3-230213A neutrino event time $T_0$. The final panel shows the residual map for the entire period from 2008 to 2025. No significant transient or steady-state \g-ray emission is observed at the KM3-230213A neutrino position. The new, unassociated excess (green circle) is still observed in the upper-right region, particularly in the 2010--11 and 2008--25 residual maps, approximately $2^\circ$ northwest from the best-fit neutrino position. The known 4FGL J0616.2-0653 source is also shown in the 2008--25 residual plot as a marginal detection, indicated by the yellow circle.}
    \label{fig:panel2}
\end{figure}

To evaluate whether electromagnetic cascades could explain the observed emission, we compare our measured SEDs with theoretical cascade spectra generated across a grid of source redshifts ($z=0.01$ -- 1.0, with 20 logarithmically spaced intervals) and IGMF strengths ($B=10^{-16}$ -- $10^{-13}$~G, with 20 logarithmically spaced intervals). For each combination of parameters, we calculate the likelihood profile by interpolating the theoretical cascade spectrum onto our measured SED points and applying an energy-dependent suppression factor accounting for pair-production-induced delays in the presence of an IGMF (Eq.~\ref{eq:approx}). Finally, we compute the 95\% CL upper limits on the $(B, z)$ parameter space by determining the flux at which the log-likelihood decreases by 2.71 from its maximum.
\begin{figure}[H]
    \centering
    \includegraphics[width=0.66\linewidth]{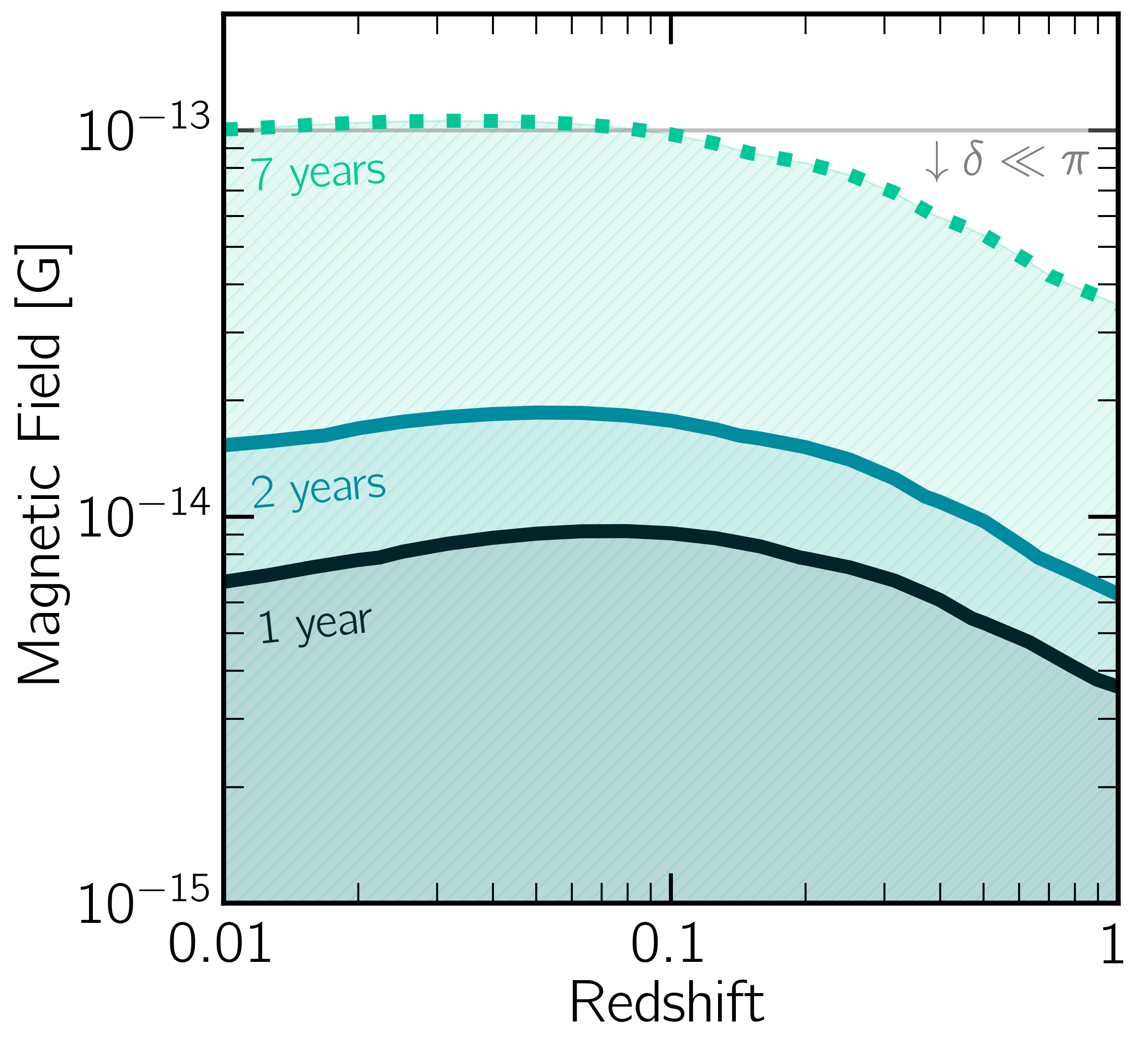}
    \caption{Constraints on the IGMF strength as a function of the redshift to the putative KM3NeT source, derived assuming IGMF evolution with $\alpha = 4$ and no detection of the cascaded $\gamma$-ray flux within periods of 1~year (black), 2~years (cyan), or 7~years (green-dashed) after the neutrino event. Shaded regions indicate parameter space excluded by our analysis. Constraints are shown at the best-fit location of the KM3NeT event, but change only negligibly over the angular uncertainty region of the event. The 1- and 2-year constraints rely on the small-angle approximation ($\delta\ll\pi$), assuming minimal \g-ray deflection. We demonstrate that extending the observation window to 7 years following the KM3-230213A (i.e., 5 more years of \textit{Fermi}-LAT observations) reaches the limit of this approximation ($\sim10^{-13}$~G).} 


    
    \label{fig:zB_limits}
\end{figure}

\section{Results}
\label{sec:analysis}
We perform a comprehensive, spatially and temporally resolved search for electromagnetic counterparts to the ultra-high-energy neutrino event KM3-230213A utilizing nearly 17 years of \FermiLAT data to constrain associated \g-ray emission predicted by the hadronic cascade model shown in Fig.~\ref{fig:main}. Our analysis involves scanning a region within $\pm2^\circ$ of the best-fit neutrino event position  in increments of 0.2$^\circ$ and examining multiple time scales (weekly, monthly, yearly, and cumulative intervals) from August 2008 to February 2025. We report no statistically significant \g-ray detection at the position of KM3-230213A during any of the analyzed time intervals (Figs.~\ref{fig:panel1} and \ref{fig:panel2}).

We specifically examine the known \FermiLAT source, \src, located approximately 0.9$^\circ$ from the best-fit neutrino event position.  Additionally, we identify a previously unknown \g-ray source located approximately $2^\circ$ from the neutrino position (still within $99\%$ KM3NeT's containment radius). We discuss both of these sources individually in the next subsections, but conclude that neither provides a convincing $\gamma$-ray counterpart to the KM3NeT signal.


Given the absence of detected transient or steady-state cascade signatures, we derive upper limits on the possible cascade-induced \g-ray flux. Comparing these limits against the theoretical cascade spectra, we constrain the potential redshift $z$ and the IGMF strength $B$. We find that incorporating the uncertainties in the EBL model reported in \cite{Saldana-Lopez:2020qzx} has a negligible impact on our results ($\lesssim 10\%$). The 95\% CL constraints are shown in Fig.~\ref{fig:zB_limits}, effectively excluding scenarios involving any magnetic fields lower than $\sim10^{-14}$~G for $z<1$ for the so-far observed data following the reported KM3NeT event.

In addition, our analysis demonstrates that extending the \textit{Fermi}‑LAT observation period by approximately 5 more years pushes us to the limit of the small-angle approximation regime used in our current cascade modeling. This extension necessitates the development of more sophisticated theoretical models to accurately describe the increased angular dispersion of cascade photons. As such, continued \textit{Fermi}‑LAT observations over the upcoming years will provide critical insight into the cascade behavior beyond the small-angle approximation, thereby allowing us to more robustly characterize the IGMF properties.


\subsection{Dedicated Analysis of 4FGL J0616.2-0653}
\label{subsec:J0616.2}
While most regions within the KM3NeT error ellipse have no detectable $\gamma$-ray emission, the source \src\ (RA = 94.06$^\circ$, Dec = -6.90$^\circ$) is a \FermiLAT \g-ray source with no established counterparts at other wavelengths. Its position within the 68\% error region of the KM3NeT neutrino event raises the question of whether the observed \g-ray excess originates from the electromagnetic cascades of a high-energy neutrino source~\cite{KM3NeT:2025bxl}.

Unlike the other regions of our analysis, the existence of a detected source means that we should not simply set an upper limit on the $\gamma$-ray luminosity, but can instead use the detected $\gamma$-ray spectrum to determine whether the $\gamma$-ray source may be produced by cascades. In particular, our models indicate that any source produced through a $\gamma$-ray cascade will have an unusually hard spectral index of $\Gamma\sim$0.9, while the vast majority of \textit{Fermi}-LAT sources have much softer spectra. Additionally, it has recently been proposed in Ref.~\cite{KM3NeT:2025bxl} that \src\ is the result of diffuse mismodeling, which also typically indicates a soft $\gamma$-ray spectrum. To investigate whether the \g-ray emission from \src\ is better explained by cascades, diffuse emission, or some other physical mechanism, we perform both spectral and temporal analyses. 

We first determine the best-fit spectral index by scanning a range of indices between $\Gamma=0.8$ and $\Gamma=4.0$ in increments of 0.1, analyzing the emission integrated over the entire \textit{Fermi}-LAT observation period (2008--2025). For each spectral index, we vary the flux normalization from $10^{-12}$ to $10^{-7}$ ph cm$^{-2}$ s$^{-1}$. The spectral analysis yields a best-fit spectral index of $\Gamma\sim2.5$, as shown in Fig.~\ref{fig:J0616}. If \src\ were a steady neutrino emitter, we would expect a persistently hard spectral emission. Thus, our results clearly disfavor an electromagnetic cascade scenario.

We then perform a finer temporal bin analysis, splitting the data into yearly intervals to search for variability or transient signals indicative of electromagnetic cascades. We find no evidence of variability trends or transient emission spatially coinciding with the KM3NeT neutrino event. We find no significant emission following KM3-230213A, in the 2023–24 time bin and, unsurprisingly, obtain a very soft spectral index beyond $\Gamma > 4.0$, suggesting a lack of any detectable hard-spectrum component. A summary of these findings is shown in Fig.~\ref{fig:J0616}.

These results indicate that the emission from \src\ is most consistent with steady diffuse emission or other soft-spectra sources (e.g., star-forming galaxies). While the persistently soft spectrum disfavors electromagnetic cascades as the dominant emission mechanism, to definitively exclude \src\ as the KM3NeT neutrino source would require a redshift measurement in order to determine whether underdeveloped cascade scenario could be relevant.
\begin{figure}[H]
    \centering
    \begin{subfigure}{0.49\textwidth}
        \centering
        \includegraphics[width=\linewidth]{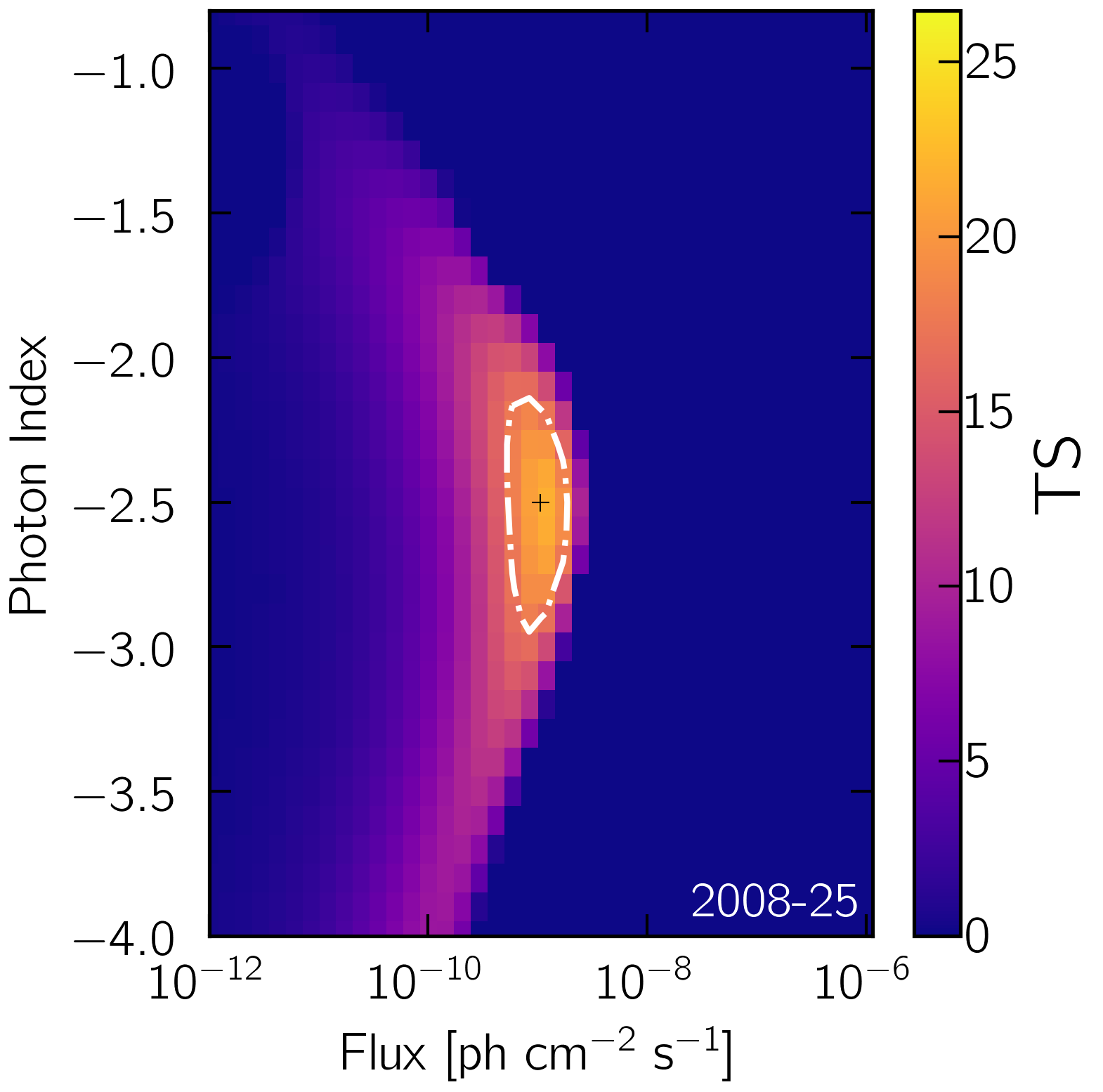}
    \end{subfigure}
    \hfill
    \begin{subfigure}{0.49\textwidth}
        \centering
        \includegraphics[width=\linewidth]{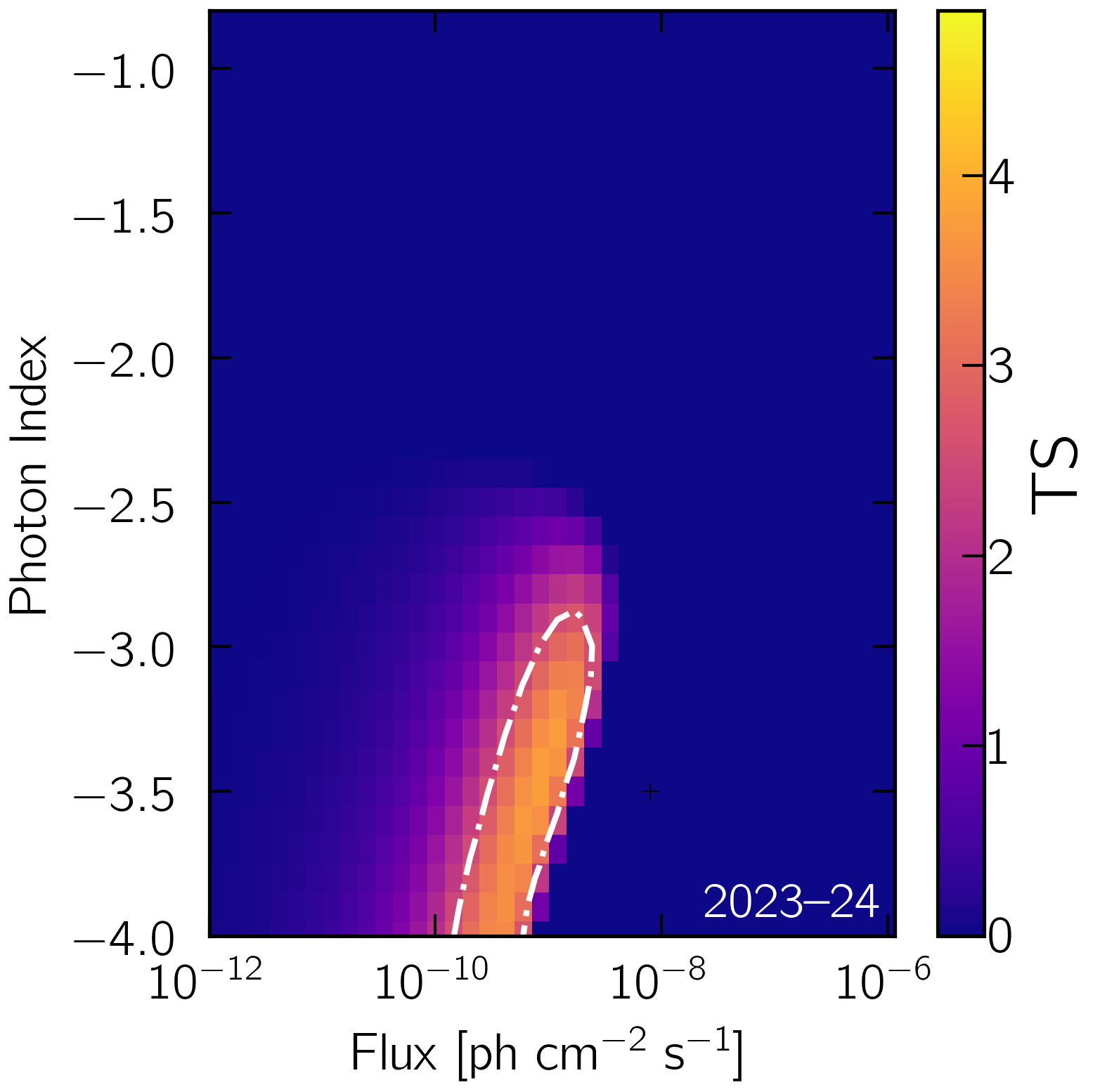}
    \end{subfigure}
    \caption{TS scan in photon index $\Gamma$ and flux for \src. \textit{Left panel}: emission integrated over the entire \textit{Fermi}-LAT observation period (2008--25), with the best-fit parameter marked by a black cross. \textit{Right panel}: emission integrated over the year following the KM3NeT neutrino event (2023--24), where the spectral index reaches the softest allowed value (no best-fit marker is shown). Dashed contours indicate the 95\% confidence area. The best-fit spectral parameters indicate a soft spectrum consistent with diffuse emission or soft-spectrum astrophysical sources rather than with the hard spectral signature expected from electromagnetic cascades.}
    \label{fig:J0616}
\end{figure}


\subsection{A New \textit{Fermi}-LAT Source near the KM3NeT Event}
\label{subsec:unassociated}

The analysis of the full \textit{Fermi}-LAT dataset reveals a \g-ray excess at a location previously unassociated with any known \g-ray source \cite{Fermi-LAT:2022byn}. The source is significantly brighter during the period between 2010--2011 than during the remainder of the analysis period. Among the candidates listed in \cite{KM3NeT:2025bxl}, the source NVSS~J060639–063421 is the closest to the identified \g-ray excess. Adopting the 17‐year best-fit localization at (RA~$= 91.97^\circ$, Dec~$= -6.47^\circ$, R(68\%) = 0.4$^\circ$) or the 2010--11 localization at (RA~$= 92.14^\circ$, Dec~$= -6.42^\circ$, R(68\%)= 0.5$^\circ$), NVSS~J060639–063421, with coordinates (RA~$\sim 91.67^\circ$, Dec~$\sim -6.57^\circ$), lies at angular distances of approximately $0.3^\circ$ and $0.5^\circ$, respectively, making it the most plausible candidate counterpart to the \g-ray excess.

NVSS~J060639–063421 is a compact radio source originally cataloged in the NVSS survey \cite{1998AJ....115.1693C}. Multiwavelength cross-identifications---in particular, detections in the radio (via VLASS, \cite{Lacy:2019rfe}), infrared (using WISE, \cite{2014yCat.2328....0C}), and tentative X-ray counterparts (eROSSITA, \cite{eROSITA:2024oyj})---suggest that NVSS~J060639–063421 may host an active galactic nucleus (AGN) with a relativistic jet, a scenario that is compatible with the production of electromagnetic cascades in hadronic acceleration models.

\begin{figure}[H]
    \centering
    \begin{subfigure}{0.49\textwidth}
        \centering
        \includegraphics[width=\linewidth]{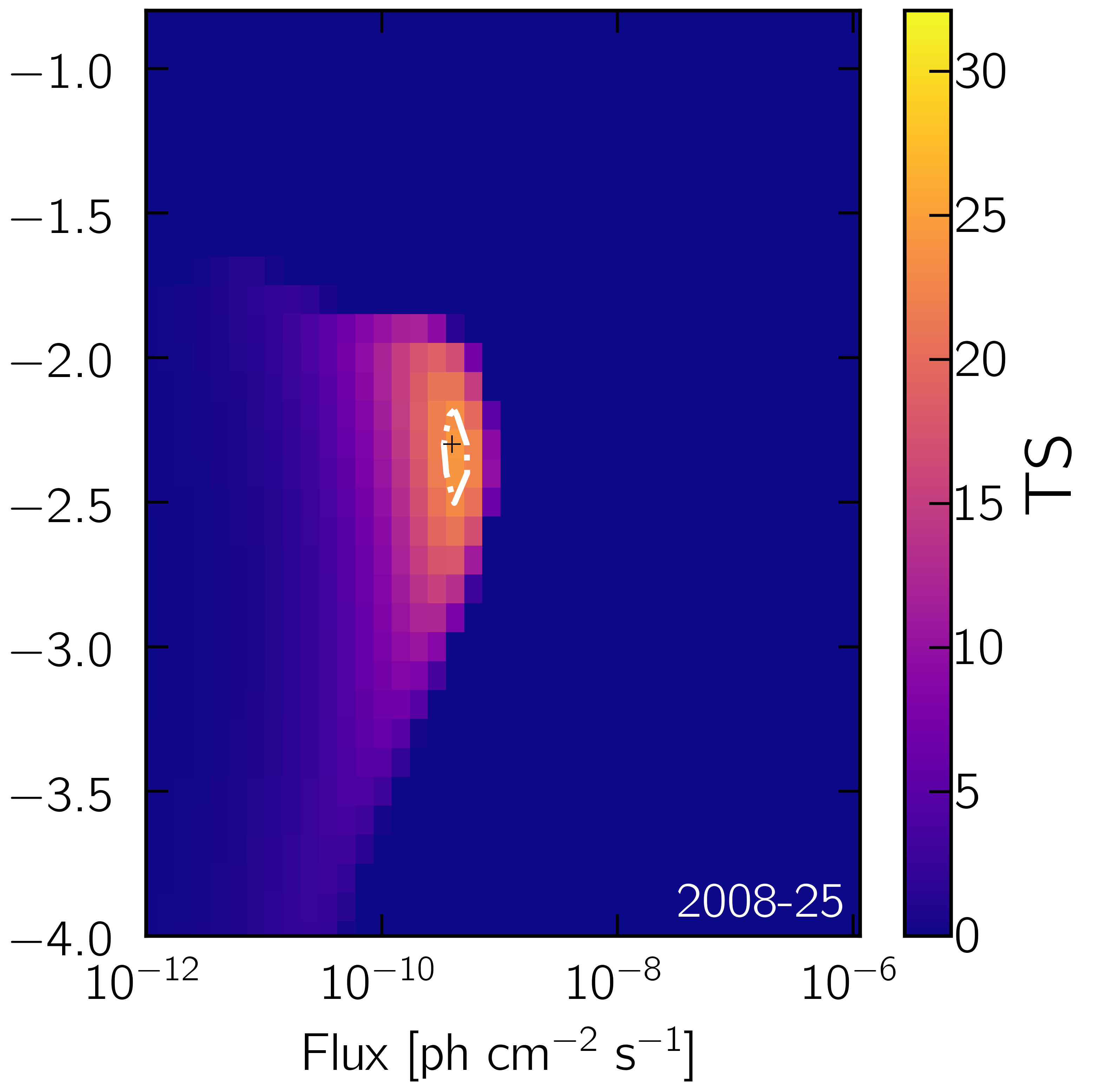}
    \end{subfigure}
    \hfill
    \begin{subfigure}{0.49\textwidth}
        \centering
        \includegraphics[width=\linewidth]{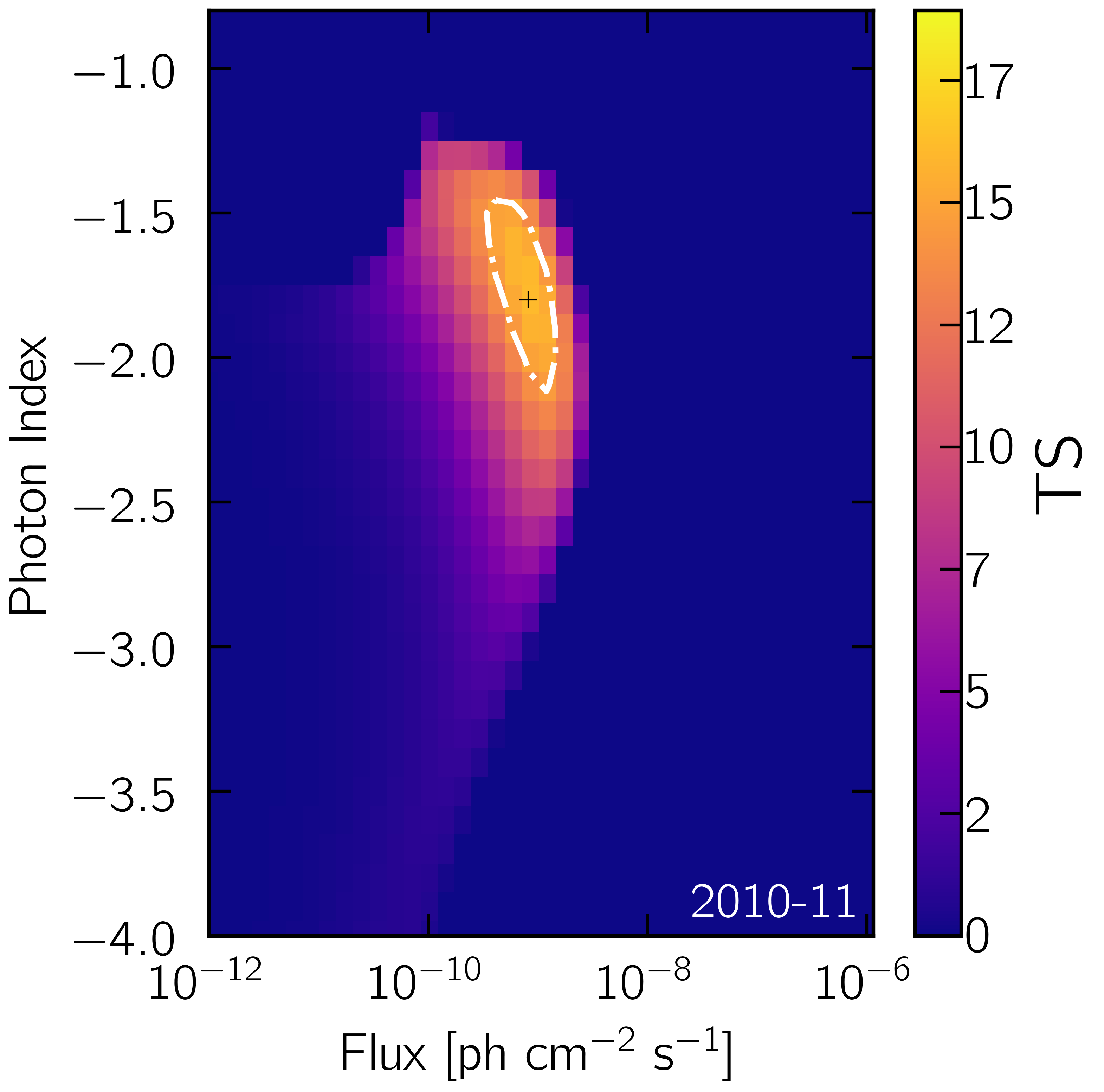}
    \end{subfigure}
        \hfill
    \begin{subfigure}{0.49\textwidth}
        \centering
        \includegraphics[width=\linewidth]{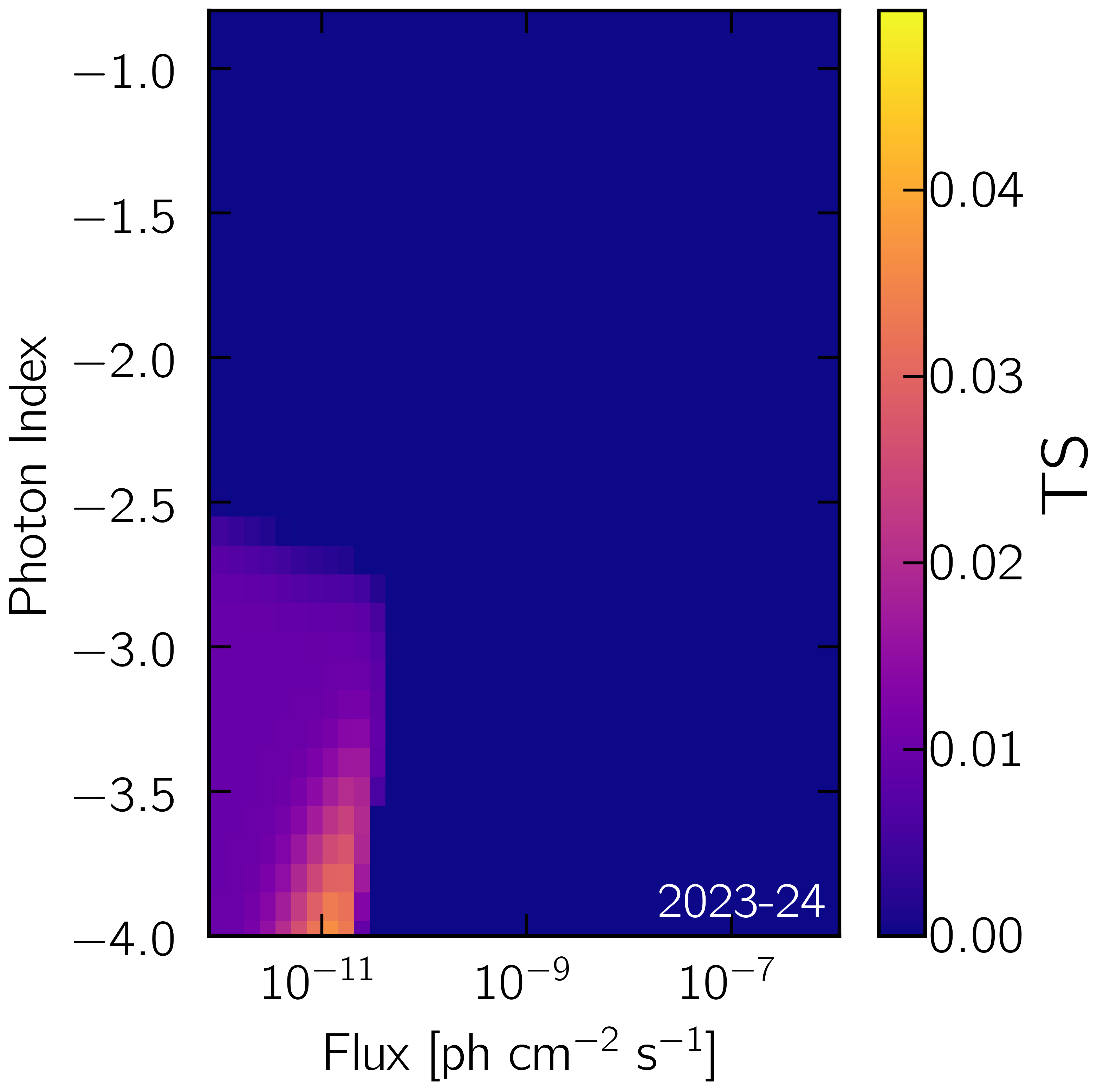}
    \end{subfigure}
    \caption{Analogous to Fig.~\ref{fig:J0616}, but for the newly identified \textit{Fermi}-LAT source located $\sim2^\circ$ from the best-fit position of KM3-230213A. We highlight a significant spectral hardening observed specifically during 2010--2011 (top right), potentially indicative of an earlier transient event consistent with low intergalactic magnetic field scenarios. The TS for the emission integrated over the year following the KM3NeT neutrino event (2023–24) is shown in the bottom row, where the spectral index reaches the softest allowed value.}
    \label{fig:new_source}
\end{figure}

To further investigate the nature of this source, we perform a time-resolved spectral analysis, splitting the dataset into yearly intervals. We find that, while the source is typically faint, it exhibits a significant spectral hardening during the 2010--2011 time frame, with a best-fit power-law index of $\Gamma \sim 1.8$, compared to its long-term average of $\Gamma \sim 2.3$ as shown in Fig.~\ref{fig:new_source}. This spectral change may suggest a possible transient episode of high-energy activity, which could be associated with a past acceleration event. While the 2010--11 spectral hardening occurred well before the KM3NeT neutrino event, making a direct temporal association unlikely, the unknown redshift of the potential counterpart NVSS~J060639--063421 prevents definitive conclusions about cascade contributions. Further constraints on the source's distance and variability, especially in multiwavelength data, is necessary to pin down the nature of this emission.

Finally, we note that our spectral analysis assumes sources at sufficient redshift ($z \gtrsim 0.1$) for fully developed electromagnetic cascades with characteristic hard spectra ($\Gamma \sim 0.9$), as predicted in \cite{Berezinsky:2016feh}. For sources at lower redshift ($z \lesssim 0.1$), the cascade development would be incomplete, producing softer spectral indices that could be consistent with the observed values we measure. However, conventional hadronic accelerators capable of producing $>$100~PeV neutrinos at very low redshift ($z \lesssim 0.01$) would need to be extraordinarily luminous. Such nearby powerful sources would be expected to have prominent counterparts across multiple wavelengths, including bright radio, X-ray, and optical emission --- which we have not yet observed for our candidate sources.

\section{Discussion}
\label{sec:discussion}

\subsection{Assumptions in Modeling the Cascade}
The simplifying model we adopt here to compute the \g-ray spectra is limited by a few assumptions. While the cascades become co-linear with the primary emission in the limit of vanishing magnetic fields, large magnetic fields can push the cascades out to several tens of degrees. In modeling the electromagnetic cascades, we assume that the angular extension of the cascade is smaller than $\sim 1^\circ$, which is true for time delays that are small enough to observe on $\mathcal{O}(1-10)$ yr time scales. We also assume a simple redshift dependence of the magnetic field, which becomes important for redshifts greater than about $z\sim$1. However, our strongest results are found at smaller redshifts, where this dependence has vanishing effects. Finally, we adopt the EBL model of Saldana-Lopez, et al. \cite{Saldana-Lopez:2020qzx}. We find that using either the $\pm 1\sigma$ fits to the EBL from this model does not significantly change our results.  

\subsection{The True Ultra-High Energy Neutrino Flux}

Throughout this paper, we assume that the KM3NeT event was produced by a putative source that not only produced one event ($k$=1), but also has a neutrino flux which was expected to produce one event ($\lambda$=1). This approach follows the recent work of Ref.~\cite{Fang:2025nzg}, which employed the same methodology to calculate the expected GeV and TeV $\gamma$-ray signals. 

This approach is reasonably justified, as the likelihood of the model expectation value of $\lambda$ given the observation of one-event ($\lambda|k$) is maximized for $\lambda$=1. However, the Poisson variance in the expected flux ($\lambda$) is large, stretching from 0.051 -- 5.57 expected events at the 2$\sigma$ level. This uncertainty propagates down to the expected $\gamma$-ray flux, which (ideally) should be normalized to match the neutrino flux, rather than the number of recorded neutrino events. In short, if KM3-230213A was a significant upward fluctuation from a relatively dim source, it is unlikely to be seen in either the \textit{Fermi}-LAT, or in TeV searches (e.g.,~\cite{Fang:2025nzg}), although the effects of this uncertainty are not explicitly modeled in Ref.~\cite{Fang:2025nzg}.

The situation is stretched farther when one first attempts to calculate the best-fit flux of the KM3NeT source that minimizes the tension between the KM3NeT detection and IceCube constraints---that is, to perform a joint-likelihood fit of KM3NeT and IceCube data. In that case, the best fit value of $\lambda$ will decrease from unity, and the corresponding uncertainty band will also decrease~\cite{Li:2025tqf}, decreasing the expected neutrino flux. The exact effect on $\lambda$ depends on the scenario used to model the neutrino source (e.g., a transient point source, steady state point source, or isotropic background). Detailed work by Ref.~\cite{Li:2025tqf} finds that the tension between the two experiments is minimized (and the best-fit value of $\lambda$ is maximized) for scenarios where KM3-230213A was produced by a transient point source, which we adopt as our default scenario.

Finally, we note that while the above discussion focused on a frequentist interpretation of the neutrino and $\gamma$-ray data, a Bayesian reconstruction treating the IceCube constraints as a prior on the flux of the KM3NeT source will give essentially equivalent results.

\section{Conclusions}
\label{sec:conclusions}
We have performed a dedicated GeV $\gamma$-ray follow-up of the ultra-high-energy neutrino event KM3-230213A using nearly 17 years of \textit{Fermi}-LAT data. Operating under the assumption that the neutrino event was produced by a transient source---a scenario that minimizes the tension between KM3NeT and IceCube observations---we find no significant \g-ray emission at the neutrino event position. Consequently, we place stringent constraints on the IGMF strength and the source refshift. Our work underscores the complementarity of neutrino and $\gamma$-ray observations, demonstrating the power of non-detections to probe new regions of astrophysical and cosmological parameter space. 


While we find no statistically significant correlations between KM3-230213A and the the GeV $\gamma$-ray events, we note two potential coincidences which required specific attention. The first pertains to the \textit{Fermi}-LAT source 4FGL J0616.2-0635. While this is a bright, and unassociated source detected in GeV $\gamma$-rays, our analysis indicates that its GeV $\gamma$-ray spectrum is quite soft, which is incompatible with models for $\gamma$-ray cascades from the KM3NeT event. Additionally, we identify a previously unknown \textit{Fermi}-LAT source, located approximately $2^\circ$ from the neutrino event (within the 99\% KM3NeT localization region). Although typically soft, this source shows significant spectral hardening during 2010--11. While this may be indicative of a prior transient event from the same source, the fact that the event is temporally resolved to only one year indicates that the IGMF must be vanishingly small, below the range constrained by Fermi-LAT blazar observations~\cite{2010Sci...328...73N}. Finally, we demonstrate that extending the \textit{Fermi}-LAT's observation period by approximately 5 more years reaches the limit of small-angle approximation regime, necessitating more sophisticated theoretical modeling, but also highlighting an important opportunity for continued \textit{Fermi}-LAT observations to more robustly characterize the IGMF properties.

One potential caveat to the above constraints stems from the possibility that the KM3NeT source is a ``hidden'' neutrino source. In this scenario, the  $\sim$400~PeV $\gamma$-rays produced alongside the neutrinos are efficiently absorbed by an intense field of low-frequency radio photons surrounding the source engine~\cite{Murase:2012df}. These ultra-high-energy photons undergo pair production through interactions with the ambient radio photons, suppressing any direct $\gamma$-ray signal, while leaving the 200~PeV neutrinos unaffected. Such a hidden-accelerator source would appear extremely bright in the radio band \cite{Fang:2025nzg}.
Furthermore, in such scenarios, the high-energy $\gamma$-ray emission is effectively absorbed and re-emitted at lower and lower energies ($\sim$MeV energies), eventually falling below the \textit{Fermi}-LAT $\gamma$-ray band. 
The detection of high energy neutrino sources without detectable GeV $\gamma$-ray counterparts thus provides a strong motivation supporting future MeV $\gamma$-ray instruments such as AMEGO \cite{AMEGO:2019gny} and e-ASTROGAM \cite{e-ASTROGAM:2016bph}.

\begin{acknowledgments}
We thank Ke Fang and Dan Hooper for helpful comments. MC and TL are supported by the Swedish Research Council under contract 2022-04283 and the Swedish National Space
Agency under contract 117/19. TL also acknowledges sabbatical support from the Wenner-Gren foundation under contract SSh2024-0037.
\end{acknowledgments}

\bibliographystyle{JHEP}
\bibliography{bibliography}

\end{document}